\documentclass[preprints,article,accept,moreauthors,pdftex]{Definitions/mdpi}
\def\slashchar#1{\setbox0=\hbox{$#1$}  
\dimen0=\wd0     
\setbox1=\hbox{/} \dimen1=\wd1  
\ifdim\dimen0>\dimen1   
\rlap{\hbox to \dimen0{\hfil/\hfil}} 
#1     
\else     
\rlap{\hbox to \dimen1{\hfil$#1$\hfil}} 
/      
\fi}
\newcommand{\dd}{\mathrm{d}}

\usepackage{amssymb}
\usepackage{comment}

\firstpage{1} 
\pubvolume{xx}
\issuenum{1}
\articlenumber{5}
\pubyear{2019}
\copyrightyear{2019}
\history{}

\Title{Parity Doubling and the Dense-Matter Phase Diagram under Constraints from Multi-Messenger Astronomy}

\Author{Michał Marczenko~$^{1,}$*\orcidA{}, 
David Blaschke~$^{1,2,3}$\orcidB{}, Krzysztof Redlich~$^{1,4}$\orcidC{} and Chihiro Sasaki $^{1}$\orcidD{}}

\AuthorNames{Michał Marczenko, David Blaschke, Krzysztof Redlich, and Chihiro Sasaki}

\address{%
$^{1}$ \quad Institute of Theoretical Physics, University of Wrocław, PL-50204 Wrocław, Poland\\
$^{2}$ \quad Bogoliubov Laboratory of Theoretical Physics, Joint Institute for Nuclear Research, 141980 Dubna, Russia\\
$^{3}$ \quad National Research Nuclear University, 115409 Moscow, Russia \\
$^{4}$ \quad Extreme Matter Institute EMMI, GSI, D-64291 Darmstadt, Germany 
}

\corres{Correspondence: michal.marczenko@ift.uni.wroc.pl}

\abstract{We extend the recently developed hybrid quark--meson--nucleon model by augmenting a six-point scalar interaction and investigate the consequences for neutron-star sequences in the mass--radius diagram. One of the characteristic features of the model is that the chiral symmetry is restored within the hadronic phase by lifting the mass splitting between chiral partner states, before quark deconfinement takes place. At low temperature and finite baryon density, the model predicts a first- or second-order chiral phase transition, or a crossover, depending on the expectation value of a scalar field, and a first-order deconfinement phase transition. We discuss two sets of free parameters, which result in compact-star mass--radius relations that are at tension with the combined constraints for maximum-mass ($2~M_\odot$) and the compactness (GW170817). We find that the most preferable mass--radius relations result in isospin-symmetric phase diagram with rather low temperature for the critical point of the chiral phase transition.}

\begin{document}

\section{Introduction}
\label{sec:introduction}
  
For the investigation of matter under extreme conditions and the structure of its phase diagram the equation of state (EoS) is the key target. In~the region of finite temperature and vanishing baryon density ab initio calculations using Monte Carlo simulations of lattice QCD~\cite{Bazavov:2014pvz} provide a benchmark for developing phenomenological approaches that can be tested, e.g.,~in heavy-ion collision experiments. Up~to now, however, the~sign problem prevents the application of lattice QCD methods to the region at low temperature and high baryon density where the possible existence of a first-order phase transition with a critical endpoint has been conjectured.   To elucidate the QCD phase structure in this domain inaccessible to terrestrial experiments and present techniques of lattice QCD simulations, valuable information comes from progress in observing the mass--radius (M-R) relationship of compact stars, due to its one-to-one correspondence with the EoS of compact star matter~\cite{Lindblom:1998dp} via the solution of the Tolman--Oppenheimer--Volkoff (TOV) equations~\cite{Tolman:1939jz,Oppenheimer:1939ne}. For~the extraction of the compact star EoS via Bayesian analysis techniques using mass and radius measurements as priors, see Refs.~\cite{Steiner:2010fz, Steiner:2012xt, Alvarez-Castillo:2016oln}. In~particular, in~the era of multi-messenger astronomy, it shall soon become possible to constrain the sequence of stable compact star configurations in the mass--radius plane inasmuch that a benchmark for the EoS of cold and dense matter can be deduced from~it. 

Among the modern observatories for measuring masses and radii of compact stars, the~gravitational wave interferometers of the LIGO-Virgo Collaboration (LVC) and the X-ray observatory NICER on-board the International Space Station provide new powerful constraints besides those from radio pulsar timing. In~this work, we pay special attention to the state-of-the-art results from the recent measurement of the high mass $2.17^{+0.11}_{-0.10}~M_\odot$ for PSR~J0740+6620 by the NANOGrav Collaboration~\cite{Cromartie:2019kug} and the compactness derived from the tidal deformability measurement for the binary compact star merger GW170817~\cite{Abbott:2018exr} in its mass range ($1.16$--$1.60~M\odot$ for the low-spin prior).

In the study of cold and dense QCD and its applications, commonly used are separate effective models for the nuclear and quark matter phases (two-phase approaches) with a priori assumed first-order phase transition, typically associated with simultaneous chiral and deconfinement transitions. Within~this setting, for~a constant-speed-of sound model of high-density (quark) matter, a~systematic classification of hybrid compact star solutions has been given in~\cite{Alford:2013aca}, which gives a possibility to identify a strong first-order transition in the EoS by the fact that the hybrid star branch in the mass--radius diagram becomes disconnected from the branch of pure neutron stars.  However, already before this occurs, a~strong phase transition manifests itself by the appearance of an almost horizontal branch on which the hybrid star solutions lie, as~opposed to the merely vertical branch of pure neutron stars. In~the literature, this strong phase transition has been discussed as due to quark deconfinement~\cite{Blaschke:2013ana,Benic:2014jia,Alvarez-Castillo:2016wqj}. This conclusion may however be premature since strong phase transitions with a large latent heat occur also within hadronic matter, for~instance due to chiral symmetry restoration within the hadronic phase~\cite{Marczenko:2018jui}. In~the present work, we employ the hybrid quark--meson--nucleon (QMN) model~\cite{Benic:2015pia, Marczenko:2017huu, Marczenko:2018jui} to explore the implications of dynamical sequential phase transitions at high baryon density on the phenomenology of neutron stars.   To improve the description of nuclear matter properties at the saturation density, we extend the previous hybrid QMN model by including a six-point scalar interaction. Our main focus is on the role of the chiral symmetry restoration within the hadronic branch of the~EoS. 
 
This paper is organized as follows. In~Section~\ref{sec:hqmn_model}, we introduce the hybrid quark--meson--nucleon model. In~Section~\ref{sec:eos}, we discuss the obtained numerical results on the equation of state under neutron-star conditions. In~Section~\ref{sec:mass_radius}, we discuss the obtained neutron-star relations. In~Section~\ref{sec:qcd_phase_diagram}, we present possible realizations of the low-temperature phase diagram. Finally, Section~\ref{sec:conclusions} is devoted to summary and~conclusions.

\section{Hybrid Quark--Meson--Nucleon~Model}
\label{sec:hqmn_model}

  In this section, we briefly introduce the hybrid QMN model for the QCD transitions at finite temperature, density, and~arbitrary isospin asymmetry for the application to the physics of neutron stars~\cite{Benic:2015pia, Marczenko:2017huu, Marczenko:2018jui}.
  
  The hybrid QMN model is composed of the baryonic parity doublet~\cite{Detar:1988kn, Jido:1999hd, Jido:2001nt} and mesons as in the Walecka model, as~well as quark degrees of freedom as in the standard quark--meson model. The~spontaneous chiral symmetry breaking yields the mass splitting between the two baryonic parity partners, while it generates an entire mass of a quark. In~this work, we consider a system with $N_f=2$; hence, relevant for this study are the lowest nucleons and their chiral partners, as~well as the up and down quarks. The~hadronic degrees of freedom are coupled to the chiral fields $\left(\sigma, \boldsymbol\pi\right)$, isosinglet vector field ($\omega_\mu$), and~isovector vector field ($\boldsymbol \rho_\mu$). The~quarks are coupled solely to the chiral fields. The~important concept of statistical confinement is realized in the hybrid QMN model by considering a medium-dependent modification of the particle distribution~functions.

  
  In the mean-field approximation, the~thermodynamic potential of the hybrid QMN model reads~\cite{Marczenko:2018jui}
\begin{equation}\label{eq:thermo_pot_iso}
     \Omega = \sum_{x}\Omega_x + V_\sigma + V_\omega + V_b + V_\rho \textrm.
  \end{equation}
  where the summation goes over the positive-parity nucleons, i.e.,~proton ($p_+$) and neutron ($n_+$), their negative-parity counterparts, denoted as $p_-$ and $n_-$, and~up~($u$) and down~($d$) quarks. The~positive-parity nucleons are identified as the positively charged and neutral $N(938)$ states. The~negative-parity states are identified as $N(1535)$~\cite{Patrignani:2016xqp}. The~kinetic part of the thermodynamic potential in Equation~\eqref{eq:thermo_pot_iso}, $\Omega_x$, reads
\begin{equation}\label{eq:thermo_potential_all}
     \Omega_x = \gamma_x \int\frac{\dd^3p}{\left(2\pi\right)^3} T \left[\ln\left(1-n_x\right) + \ln\left(1-\bar n_x\right)\right]\textrm.
  \end{equation}
  
  The spin degeneracy of the nucleons is $\gamma_\pm=2$ for both positive- and negative-parity states, while the color-spin degeneracy factor for quarks is $\gamma_q=2\times 3 = 6$. The~functions $n_x$ are the modified \mbox{Fermi--Dirac} distribution functions for the nucleons
\begin{equation}\label{eq:cutoff_nuc}
  \begin{split}
      n_\pm      &= \theta \left(\alpha^2 b^2 - \boldsymbol p^2\right) f_\pm \textrm,\\
      \bar n_\pm &= \theta \left(\alpha^2 b^2 - \boldsymbol p^2\right) \bar f_\pm \textrm,
  \end{split}
  \end{equation}
  and for the quarks, accordingly
\begin{equation}\label{eq:cutoff_quark}
  \begin{split}
     n_q      &= \theta \left(\boldsymbol p^2-b^2\right) f_q \textrm,\\
     \bar n_q &= \theta \left(\boldsymbol p^2-b^2\right) \bar f_q \textrm,
  \end{split}
  \end{equation}
  where $b$ is the expectation value of the $b$-field, and~$\alpha$ is a dimensionless model parameter~\cite{Benic:2015pia, Marczenko:2017huu}.

  The hybrid QMN model employs confinement/deconfinement mechanism in a statistical sense. The~approach used in this model is to introduce IR and UV momentum cutoffs to suppress quarks at low momenta and hadrons at high momenta. This notion has been widely used in effective theories and Dyson--Schwinger approaches~\cite{Roberts:2010rn, Roberts:2011wy}. In~the current approach, the~cutoff is replaced with a medium-dependent quantity, which is expected from asymptotic freedom. Such an intrinsic modification of the cutoff is determined self-consistently when the cutoff is regarded as a vacuum expectation value of a scalar field (see Equation~\eqref{eq:potentials_b}). The~role of the $\alpha b_0$ parameter can be understood twofold. First, its lower values trigger the chiral phase transition at lower densities. Second, the~chiral phase transition is stronger and the equation of state becomes stiffer for lower values of the parameter. This can be seen in the equation of state and corresponding speed of sound squared as functions of net-baryon-number density (see Section~\ref{sec:eos}).

 The functions $f_x$ and $\bar f_x$ are the standard Fermi--Dirac distributions,
\begin{equation}
  \begin{split}
    f_x      &= \frac{1}{1+e^{\beta \left(E_x - \mu_x\right)}} \textrm,\\
    \bar f_x &= \frac{1}{1+e^{\beta \left(E_x + \mu_x\right)}}\textrm,
  \end{split}
  \end{equation}
  with $\beta$ being the inverse temperature, the~dispersion relation $E_x = \sqrt{\boldsymbol p^2 + m_x^2}$. The~effective chemical potentials for $p_\pm$ and $n_\pm$ are  {defined }as\footnote{In the mean-field approximation, the~non-vanishing expectation value of the $\omega$ field is the time-like component; hence, we simply denote it by $\omega_0 \equiv \omega$. Similarly, we denote the non-vanishing component of the $\rho$ field, time-like and neutral, by~$\rho_{03} \equiv \rho $.}  
\begin{equation}\label{eq:u_eff_had_iso}
  \begin{split}
    \mu_{p_\pm} &= \mu_B - g_\omega\omega - \frac{1}{2}g_\rho \rho + \mu_Q\textrm,\\
    \mu_{n_\pm} &= \mu_B - g_\omega\omega + \frac{1}{2}g_\rho \rho\textrm.
  \end{split}
  \end{equation}
  
  The effective chemical potentials for up and down quarks are given by
\begin{equation}\label{eq:u_effq}
  \begin{split}
    \mu_u &= \frac{1}{3}\mu_B + \frac{2}{3}\mu_Q\textrm,\\
    \mu_d &= \frac{1}{3}\mu_B - \frac{1}{3}\mu_Q\textrm.
  \end{split}
  \end{equation}
  
  In Equations~(\ref{eq:u_eff_had_iso})~and~(\ref{eq:u_effq}), $\mu_B$ and $\mu_Q$ are the baryon and charge chemical potentials, respectively. The~constants $g_\omega$ and $g_\rho$ couple the nucleons to the $\omega$ and $\rho$ fields, respectively. The~strength of $g_\omega$ is fixed by the nuclear saturation properties, while the value of $g_\rho$ can be fixed by fitting the value of symmetry energy~\cite{Lattimer:2012xj}. The~properties are shown in Table~\ref{tab:external_params}. 
 
 \begin{table}[H]
\centering 
    \caption{Properties of the nuclear ground state at $\mu_B = 923~$MeV and the symmetry energy used in this~work.}
    \label{tab:external_params}
  \begin{tabular}{cccc}
    \toprule
    \boldmath{$\rho_0~$}\textbf{[fm}\boldmath{$^{-3}$}\textbf{]} & \boldmath{$E/A - m_+$} \textbf{[MeV]} & \boldmath{$K$}~\textbf{[MeV]} & \boldmath{$E_{\rm sym}$} \textbf{[MeV]} \\ \midrule
    $0.16$               & $-16$             & 240       & 31 \\ \bottomrule
    \end{tabular}
  \end{table}

The effective masses of the parity doublers \mbox{$m_{p_\pm} = m_{n_\pm} \equiv m_\pm$} are given by
\begin{equation}\label{eq:mass_had}
    m_\pm = \frac{1}{2}\left[\sqrt{\left(g_1+g_2\right)^2\sigma^2 + 4m_0^2} \mp \left( g_1 - g_2 \right) \sigma \right] \textrm,
  \end{equation}
  and for quarks, $m_u = m_d \equiv m_q$,
\begin{equation}\label{eq:mass_quark}
    m_q = g_q \sigma \textrm.
  \end{equation}
  
  The parameters $g_1$, $g_2$, and $g_q$ are Yukawa-coupling constants, $m_0$ is the chirally invariant mass of the baryons and is treated as an external parameter (for more details, see  \cite{Marczenko:2017huu, Marczenko:2018jui}). The~values of those couplings can be determined by fixing the fermion masses in the vacuum (see Table~\ref{tab:vacuum_params}). The~quark mass is assumed to be $m_+ = 3 m_q$ in the vacuum. When the chiral symmetry is restored, the~masses of the baryonic parity partners become degenerate with a common finite mass $m_\pm\left(\sigma=0\right) = m_0$, which reflects the parity doubling structure of the \mbox{low-lying} baryons. This is in contrast to the quarks, which become massless as the chiral symmetry gets~restored.

  \begin{table}[H]
   \caption{Physical vacuum inputs used in this~work.}
    \label{tab:vacuum_params}
\centering
  \begin{tabular}{cccccc}
    \toprule
    \boldmath{$m_+~$}\textbf{[MeV]} & \boldmath{$m_-~$}\textbf{[MeV]} & \boldmath{$m_\pi~$}\textbf{[MeV]} & \boldmath{$f_\pi~$}\textbf{[MeV]} & \boldmath{$m_\omega~$}\textbf{[MeV]} & \boldmath{$m_\rho~$}\textbf{[MeV]} \\ \midrule
    939   & 1500  & 140     & 93      & 783        & 775     \\ \bottomrule
    \end{tabular}
    \end{table}

    The potentials in Equation~\eqref{eq:thermo_pot_iso} are as in the SU(2) linear sigma model,
  \begin{subequations}\label{eq:potentials}
\begin{align}
    V_\sigma &= -\frac{\lambda_2}{2}\left(\sigma^2 + \boldsymbol\pi^2\right) + \frac{\lambda_4}{4}\left(\sigma^2 + \boldsymbol\pi^2\right)^2 - \frac{\lambda_6}{6}\left(\sigma^2 + \boldsymbol\pi^2\right)^3- \epsilon\sigma \textrm,\label{eq:potentials_sigma}\\
    V_\omega &= -\frac{1}{2}m_\omega^2 \omega_\mu\omega^\mu\textrm,\\
    V_b &= -\frac{1}{2} \kappa_b^2 b^2 + \frac{1}{4}\lambda_b b^4 \textrm,\\
    V_\rho &= - \frac{1}{2}m_\rho^2{\boldsymbol \rho}_\mu{\boldsymbol \rho}^\mu \textrm,\label{eq:potentials_b}
  \end{align}
  \end{subequations}
  where $\lambda_2 = \lambda_4f_\pi^2 - \lambda_6f_\pi^4 - m_\pi^2$, and~$\epsilon = m_\pi^2 f_\pi$. $m_\pi$, $m_\omega$, and~$m_\rho$ are the $\pi$, $\omega$, and~$\rho$ meson masses, respectively, The pion decay constant is denoted as $f_\pi$. Their values are shown in Table~\ref{tab:vacuum_params}. The~constants $\kappa_b$ and $\lambda_b$ are fixed following Ref.~\cite{Benic:2015pia}. The~parameters $\lambda_4$ and $\lambda_6$ are fixed by the properties of the nuclear ground state (see Table~\ref{tab:external_params}). We note that the introduction of the six-point scalar interaction term in Equation~\eqref{eq:potentials_sigma} is essential in order to reproduce the experimental value of the compressibility \mbox{$K=240\pm20~$MeV~\cite{Motohiro:2015taa}}.

  Following the previous studies of the \mbox{parity-doublet-based} models~\cite{Zschiesche:2006zj, Benic:2015pia, Marczenko:2017huu, Marczenko:2018jui}, as~well as recent lattice QCD results~\cite{Aarts:2017rrl, Aarts:2018glk}, we choose rather large values, $m_0=700,~800$~MeV. We note that the additional mass, $m_0$, is not associated with spontaneous chiral symmetry breaking. Thus, it has to originate through another mechanism. Although~it is unknown how $m_0$ is expressed in terms of the QCD condensates, the~constraint $m_0 \leq 800~$MeV is transmuted into the nucleon mass such that at most $15\%$ of the entire mass is generated by the spontaneous chiral symmetry breaking. This is best seen in the chiral limit, where no dimensionful parameters are present in the QCD Lagrangian, but~the appearance of the QCD scale breaks the scale invariance. Thus, one expects that both give rise to the emergence of dynamical hadronic scales at low energies~\cite{Collins:1976yq, Bardeen:1985sm, Nielsen:1977sy}. Thus, the~chirally invariant mass, $m_0$, can be identified with the gluon condensate $\langle G_{\mu\nu}G^{\mu\nu} \rangle$.

  The physical inputs and the model parameters used in this work are summarized in Tables~\ref{tab:external_params}--\ref{tab:model_params}. In-medium profiles of the mean fields are obtained by extremizing the thermodynamic potential~in Equation (\ref{eq:thermo_pot_iso}). The~gap equations are obtained as~follows
  
  \begin{subequations}\label{eq:gap_eqs_iso}
\begin{align}
    \frac{\partial\Omega}{\partial\sigma} &= -\lambda_2\sigma + \lambda_4\sigma^3 -\lambda_6\sigma^5 - \epsilon + \sum_{x=p_\pm,n_\pm, u,d}s_x \frac{\partial m_x}{\partial \sigma} = 0 \textrm, \label{eq:gap_eq_sigma}\\
    \frac{\partial\Omega}{\partial\omega} &= -m_\omega^2 \omega + g_\omega\sum_{x=p_\pm,n_\pm}\rho_x = 0 \textrm,\label{eq:gap_omega}\\
    \frac{\partial\Omega}{\partial b} &= - \kappa_b^2 b + \lambda_b b^3 + \alpha \sum_{x=p_\pm,n_\pm} \hat\omega_x - \sum_{x=u,d}\hat\omega_x = 0 \textrm,\label{eq:gap_b}\\
    \frac{\partial\Omega}{\partial\rho} &= -m_\rho^2 \rho + \frac{1}{2}g_\rho\sum_{x=p_\pm}\rho_x - \frac{1}{2}g_\rho\sum_{x=n_\pm}\rho_x = 0 \textrm,
  \end{align}
  \end{subequations}
  where the scalar and baryon densities are
\begin{equation}\label{eq:scalar_den}
    s_x = \gamma_x \int\frac{\dd^3 p}{(2\pi)^3}\; \frac{m_x}{E_x} \left( n_x + \bar n_x \right) \textrm,
  \end{equation}
  and
\begin{equation}\label{eq:vector_den}
    \rho_x = \gamma_x \int\frac{\dd^3 p}{(2\pi)^3}\; \left( n_x - \bar n_x \right) \textrm,
  \end{equation}
  respectively. The~boundary terms in the gap Equation~\eqref{eq:gap_b} are given as
\begin{equation}\label{eq:boundary_nucleon}
    \hat\omega_{\pm} = \gamma_{\pm} \frac{(\alpha b)^2}{2\pi^2} T \left[ \ln\left(1 - f_{\pm}\right) + \ln\left(1 - \bar f_{\pm}\right) \right]_{\boldsymbol p^2 = (\alpha b)^2}\textrm,
  \end{equation}
  and
\begin{equation}\label{eq:boundary_quark}
    \hat\omega_q = \gamma_q \frac{b^2}{2\pi^2} T \left[ \ln\left(1 - f_q\right) + \ln\left(1 - \bar f_q\right) \right]_{\boldsymbol p^2 = b^2} \textrm,
  \end{equation}
  for the nucleons and quarks, respectively. Note that the terms in Equations (\ref{eq:boundary_nucleon}) and (\ref{eq:boundary_quark}) come into the gap Equation~(\ref{eq:gap_b}) with opposite signs. This reflects the fact that nucleons and quarks favor different values of the bag~field.

\unskip

\unskip

  \begin{table}[H]\centering
    \caption{Sets of the model parameters used in this work. The~values of $\lambda_4$, $\lambda_6$ and $g_\omega$ are fixed by the nuclear ground state properties, and~$g_\rho$ by the symmetry energy (see the text). The~remaining parameters, $g_q$, $\kappa_b$, and~$\lambda_b$, do not depend on the choice of $m_0$, and~their values are taken from Ref.~\cite{Marczenko:2017huu}.}
    \label{tab:model_params}
  \begin{tabular}{cccccccccc}
    \toprule
    \boldmath{$m_0$}~\textbf{[MeV]} & \boldmath{$\lambda_4$} & \boldmath{$\lambda_6f_\pi^2$} & \boldmath{$g_\omega$} &  \boldmath{$g_\rho$} & \boldmath{$g_1$} & \boldmath{$g_2$} & \boldmath{$g_q$} & \boldmath{$\kappa_b~$}\textbf{[MeV]} & \boldmath{$\lambda_b$} \\ \midrule
    700         & 33.74          & 13.20           & 5.60       & 8.10      & 13.75 & 7.72 & \multirow{2}{*}{3.36} & \multirow{2}{*}{155} & \multirow{2}{*}{0.074} \\ \cmidrule{1-7}
    800         & 21.50           & 8.25           & 7.27       & 7.92      & 12.91 & 6.88 & \multirow{2}{*}{} & \multirow{2}{*}{} & \multirow{2}{*}{} \\ \bottomrule
  \end{tabular}
  
  \end{table}
   
  In the grand canonical ensemble, the~thermodynamic pressure is obtained from the thermodynamic potential as $P = -\Omega + \Omega_0$, where $\Omega_0$ is the value of the thermodynamic potential in the vacuum. The~net-baryon number density for a species $x$ is defined as
\begin{equation}
    \rho^x_B = -\frac{\partial \Omega^x}{\partial \mu_B} \textrm,
  \end{equation}
  where $\Omega^x$ is the kinetic term in Equation~\eqref{eq:thermo_potential_all} for the species $x$. The~total net-baryon number density~reads
\begin{equation}
    \rho_B = \rho_B^{n_+} + \rho_B^{n_-} + \rho_B^{p_+} + \rho_B^{p_-} + \rho_B^{u} + \rho_B^{d} \textrm.
  \end{equation}

  In the next section, we discuss the   obtained equations of state in the hybrid QMN model and their impact on the chiral phase transition, under~the neutron-star conditions of $\beta$ equilibrium and charge~neutrality.
   
\section{Equation of State under Neutron-Star~Conditions}
\label{sec:eos}

  The neutron-star conditions require additional constraints to be imposed on the EoS under investigation. To~this end, electrons and muons are included as gases of free relativistic particles. The first constraint is the $\beta$-equilibrium. This condition is an equilibrium  among  protons, neutrons, and~charged leptons. It assumes that the energy of the system is minimized, the~system is electrically neutral, and~the total net-baryon number is conserved. $\beta$-equilibrium condition can be expressed in terms of chemical potentials,
\begin{equation}
    \mu_{n_+} = \mu_{p_+} + \mu_{e/\mu} \textrm,
  \end{equation}
  where $\mu_{n_+}$, $\mu_{p_+}$, $\mu_{e}$, and $\mu_\mu$ are the neutron, proton, electron, and muon chemical potentials, respectively. The~electric-charge neutrality constraint dictates that the overall charge density in a neutron star has to be zero,
\begin{equation}
    \rho_Q^{p_+} + \rho_Q^{p_-} + \frac{2}{3}\rho_Q^u - \frac{1}{3}\rho_Q^d - \rho_Q^e - \rho_Q^\mu = 0 \textrm,
  \end{equation}
  where $\rho_Q^x$ is the charge density of a species $x$.

\begin{figure}[H]
\centering
    \includegraphics[width=0.497\linewidth]{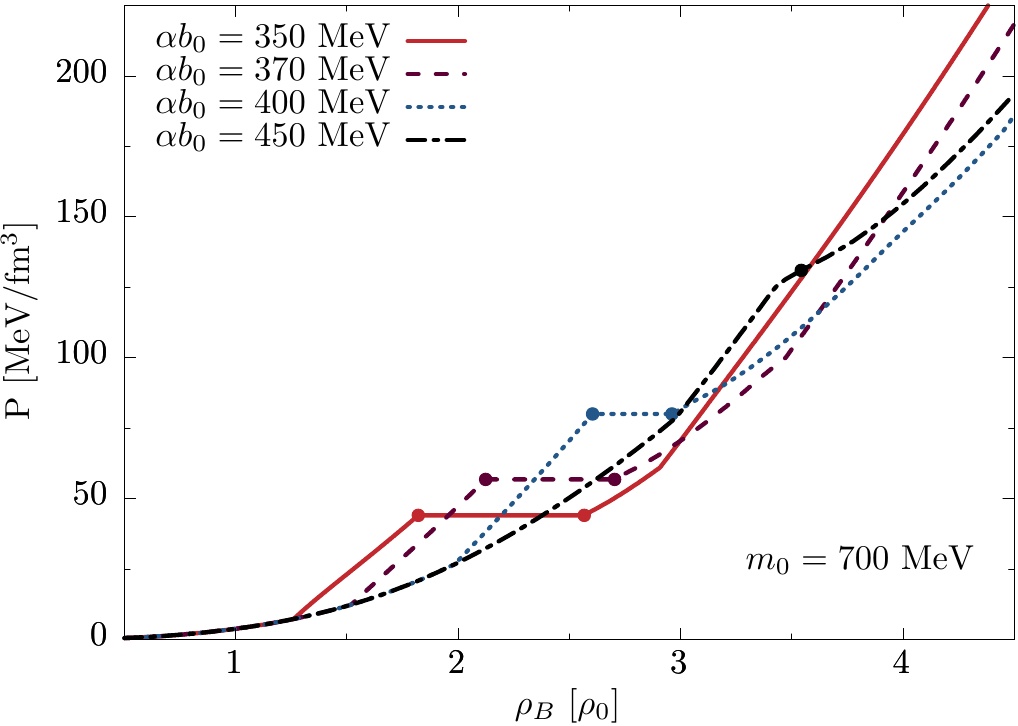}
    \includegraphics[width=0.497\linewidth]{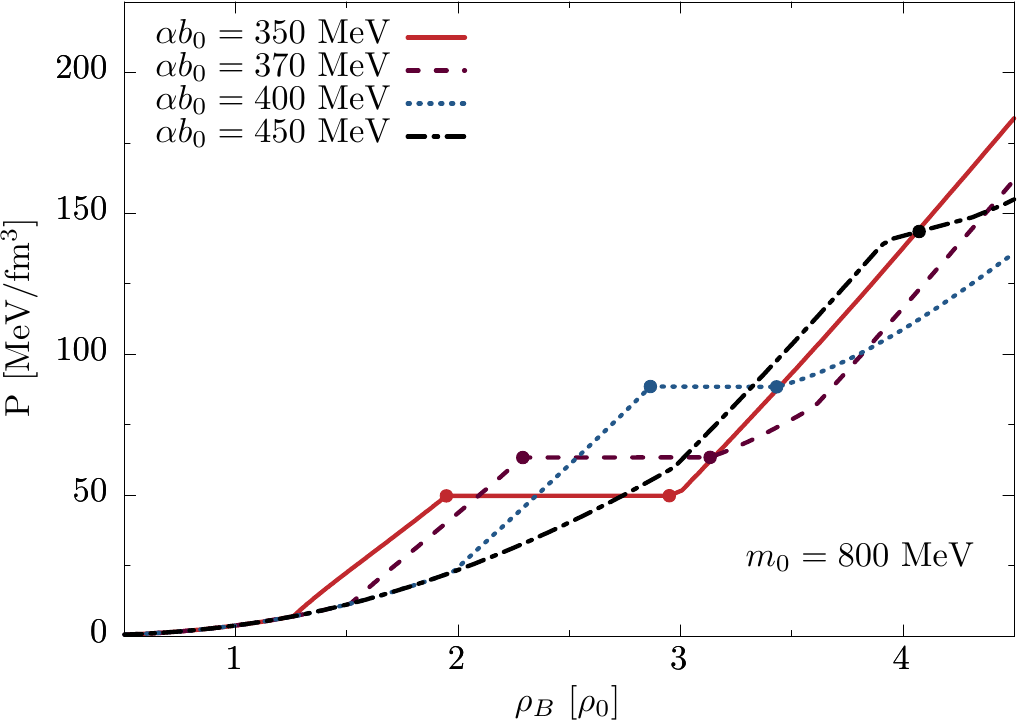}
    \caption{Thermodynamic pressure $P$ as a function of the net-baryon number density $\rho_B$, in~units of the saturation density, $\rho_0=0.16$~fm$^{-3}$ for $m_0=700~$MeV (\textbf{left}) and $m_0=800~$MeV (\textbf{right}). The~regions between circles correspond to the coexistence of chirally broken and restored phases in the first-order phase transition. For~\mbox{$\alpha b_0=450~$MeV}, the transition is a crossover. The~deconfinement transitions are triggered at higher densities and are not shown~here.}
    \label{fig:eos} 
  \end{figure}

  In Figure~\ref{fig:eos}, we show the calculated zero-temperature equations of state in the mean-field approximation with $m_0=700~$MeV (Figure~\ref{fig:eos}, left) and $m_0=800~$MeV (Figure~\ref{fig:eos}, right), for~different values of the $\alpha$ parameter, namely \mbox{$\alpha b_0=350$~MeV} (red, solid line), \mbox{$\alpha b_0=370$~MeV} (purple, dashed line), \mbox{$\alpha b_0=400$~MeV} (blue, dotted line) and \mbox{$\alpha b_0=450$~MeV} (black, dash-dotted line). The~value $b_0$ denotes the vacuum expectation value of the $b$-field. The~coexistence phases of the chirally broken and restored phases are shown between circles. We stress that the chiral and deconfinement phase transitions are sequential in the current model setup (see  \cite{Marczenko:2017huu}). The~latter happen at higher densities and are not shown in the~figure.

  In all cases, the~behavior at low densities is similar. In~general, for~low values of $\alpha b_0$ (except $\alpha b_0=450~$MeV), the~chiral transition is of first order, determined as a jump in the $\sigma$-field expectation value. The~parity-doublet nucleons become degenerate with mass $m_\pm = m_0$. The~chiral phase transition becomes weaker for higher values of the $\alpha$ parameter. For~$\alpha b_0=450$~MeV, the~transition turns into a smooth crossover, defined as a peak in $\partial \sigma / \partial \mu_B$. This behavior agrees with the case of isospin-symmetric matter, where higher value of $\alpha$ causes the first-order chiral phase transition to weaken and eventually go through a critical point, and~turn into a crossover transition~\cite{Marczenko:2017huu}. The~values of the net-baryon density range for the coexistence phase of the chirally broken and restored phases are shown in Table~\ref{tab:chiral_transition}.

\begin{table}[H]\centering
   \caption{Net-baryon density range of the coexistence phase of the chirally broken and restored phases in terms of saturation density units, $\rho_0$, for~different values of $m_0$ and $\alpha b_0$ parameters. For~the case of $\alpha b_0=450~$MeV, the~transitions are smooth crossovers for both values of $m_0$.}
    \label{tab:chiral_transition}
  \begin{tabular}{ccccc}
    \toprule
                & \multicolumn{4}{c}{\boldmath{$\alpha b_0~$} \textbf{[MeV]}}         \\ \midrule
    \boldmath{$m_0$}~\textbf{[MeV]} & \textbf{350}       & \textbf{370}       & \textbf{400}       & \textbf{450}  \\ \midrule
    700         & $1.82$--$2.60$ 
    & $2.12$--$2.76$ & $2.60$--$3.07$ & $3.56$ \\ \midrule
    800         & $1.94$--$2.97$ & $2.29$--$3.15$ & $2.86$--$3.66$ & $4.13$ \\ \bottomrule
    \end{tabular}

  \end{table}
  
\begin{figure}[H]
\centering
    \includegraphics[width=0.497\linewidth]{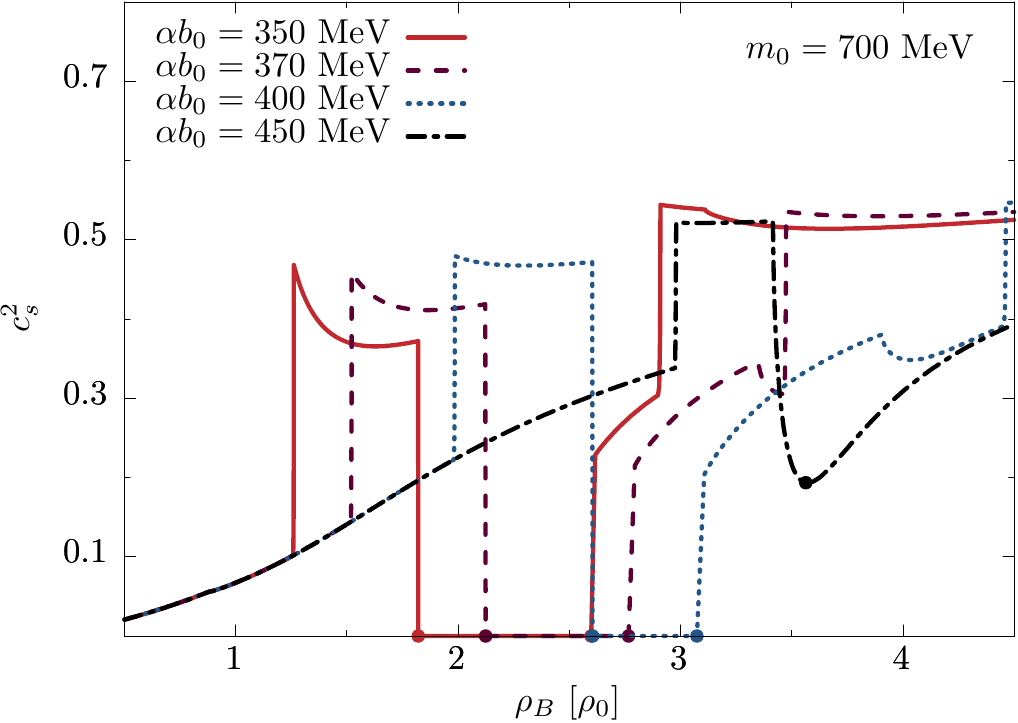}
    \includegraphics[width=0.497\linewidth]{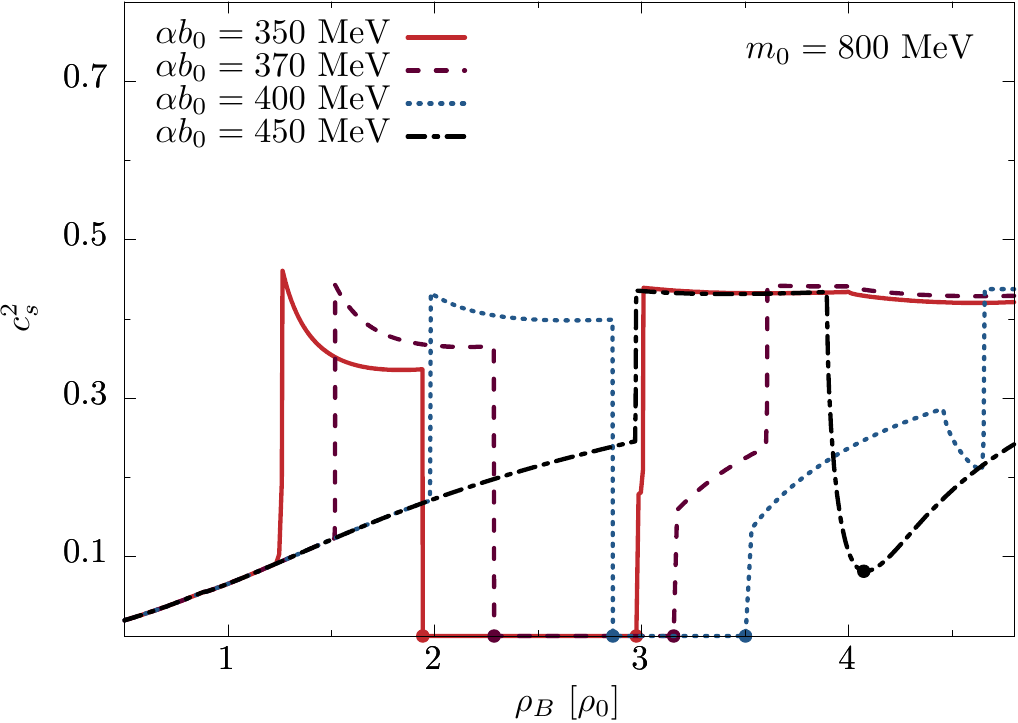}
    \caption{Speed of sound squared as a function of the energy density for $m_0=700~$MeV (\textbf{left}) and $m_0=800~$MeV (\textbf{right}). The~regions between circles correspond to the coexistence of chirally broken and restored phases in the first-order phase transition. For~\mbox{$\alpha b_0=450~$MeV}, the transition is a crossover. The~deconfinement transitions are triggered at higher densities and are not shown~here.}
    \label{fig:cs}
  \end{figure}

    In Figure~\ref{fig:cs}, we show the speed of sound squared, $c_s^2 = \dd P / \dd \epsilon$, in~units of the speed of light squared, as~a function of the net-baryon number density. The~coexistence phases are shown in between circles. As~seen in the figure, the~causality bound is preserved for all of the parameterizations. The~apparent stiffening of the EoSs is a result of the modification of the Fermi--Dirac distributions (cf.~Equation~\eqref{eq:cutoff_nuc}) introduced in the hybrid QMN model. We note that it is in general possible to sustain the $2~M_\odot$ constraint and fulfill the conformal bound, i.e.,~$c_s^2 \leq 1/3$. This can be obtained, e.g.,~in a class of constant-speed-of-sound equations of state~\cite{Alford:2015dpa}.

\unskip

\unskip

\section{TOV Solutions for Compact-Star~Sequences}
\label{sec:mass_radius}

  We use the equations of state introduced in the previous section (see Figure~\ref{fig:eos}) to solve the general-relativistic TOV equations~\cite{Tolman:1939jz, Oppenheimer:1939ne} for spherically symmetric objects, 
\vspace{12pt}
  \begin{subequations}\label{eq:TOV_eqs}
\begin{align}
     \frac{\dd P(r)}{\dd r} &= -\frac{\left(\epsilon(r) + P(r)\right)\left(M(r) + 4\pi r^3 P(r)\right)}{r \left(r-2M(r)\right)} \textrm,\\
     \frac{\dd M(r)}{\dd r} &= 4\pi r^2 \epsilon(r)\textrm,
  \end{align}
  \end{subequations}
  with the boundary conditions \mbox{$P(r=R) = 0$} and  \mbox{$M = M(r=R)$}, where $R$ and $M$ are the radius and the mass of a neutron star, respectively. Once the initial conditions are specified based on a given equation of state, namely the central pressure $P_c$ and the central energy density $\epsilon_c$, the~internal profile of a neutron star can be~calculated.

\begin{figure}[H]
 \centering
    \includegraphics[width=1\linewidth]{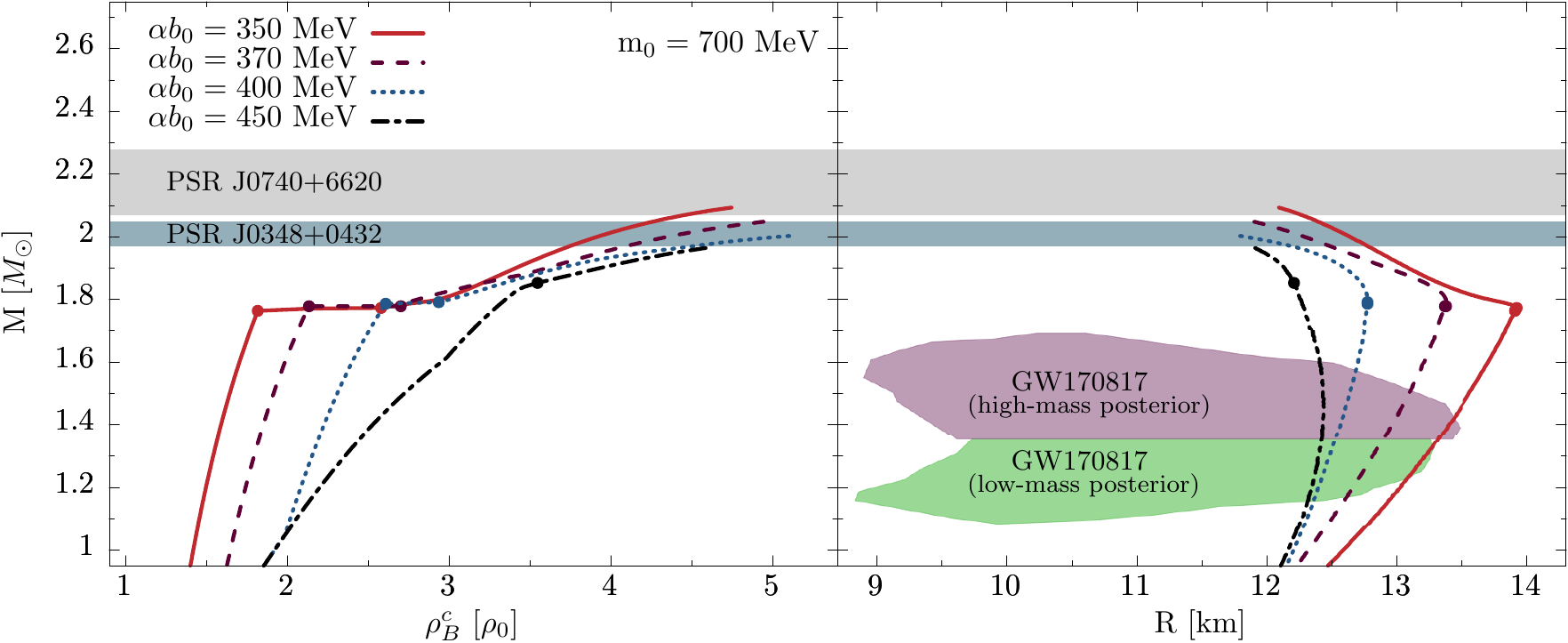}\\\vspace{10pt}
    \includegraphics[width=1\linewidth]{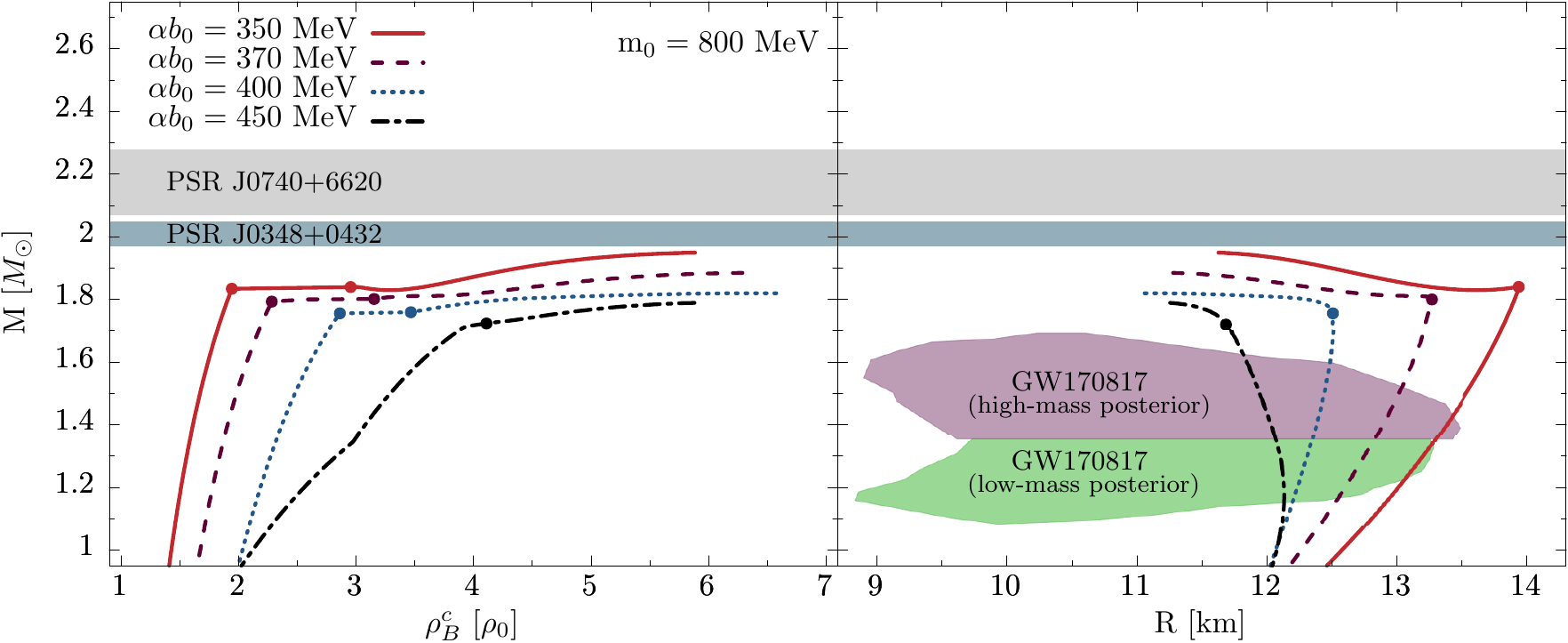}
    \caption{Sequences of mass for compact stars vs. their central net-baryon density (\textbf{left}) and vs. radius (\textbf{right}) as solutions of the TOV equations for $m_0=700$~MeV (\textbf{top}) and $m_0=800$~MeV (\textbf{bottom}). The~regions between the circles show the coexistence of the chirally broken and chirally restored phases. The~gray band shows the $2.17^{+0.11}_{-0.10}~M_\odot$ constraint~\cite{Cromartie:2019kug}. The~blue band is the $2.01\pm0.04~M_\odot$ constraint~\cite{Antoniadis:2013pzd}. The~green and purple bands in the right panel show $90\%$ credibility regions obtained from the GW170817 event~\cite{Abbott:2018exr} for the low- and high-mass~posteriors.}
    \label{fig:m_panel}
  \end{figure}

  In general, there is one-to-one correspondence between an EoS and the \mbox{mass--radius} relation calculated with it. In~  Figure~\ref{fig:m_panel} (left), we show the relationship of mass vs. central net-baryon number density, for~the calculated sequences of compact stars, together with the state-of-the-art constraints on the maximum mass for the pulsar PSR~J0348{-0432}~\cite{Antoniadis:2013pzd} and PSR~J0740+6620~\cite{Cromartie:2019kug}. We point out that the chiral phase transition leads to a softening of the EoS so that it is accompanied by a rapid flattening of the sequence. Notably, the~chiral transition for all values of $\alpha b_0$ occurs in the \mbox{high-mass} part of the sequence, but~below the \mbox{$2~M_\odot$} constraint, at~around $1.8~M_\odot$.

In Figure~\ref{fig:m_panel} (left), the~three curves for \mbox{$\alpha b_0 = 350,~370,~400~$MeV} consist of three phases: the chirally broken phase in the low-mass part of the sequence, the~chirally restored phase in the \mbox{high-mass} part, and~the coexistence phase between filled circles. Similar  to the equation of state, increasing the value of $\alpha$ softens the chiral transition, which eventually becomes a smooth crossover for $\alpha b_0 = 450~$MeV and consists only of branches with chiral symmetry being broken and restored, separated by a~circle.

\begin{figure}[H]
  \centering
    \includegraphics[width=0.497\linewidth]{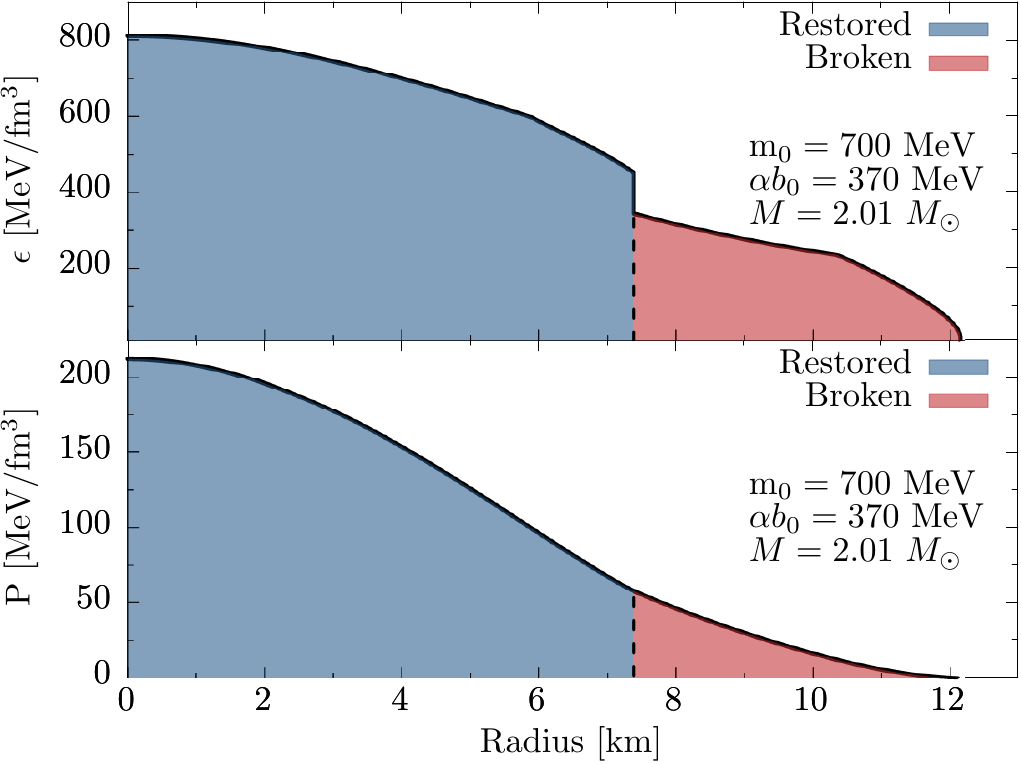}
    \caption{Profiles of the energy density (\textbf{top}) and pressure (\textbf{bottom}) for a neutron star with $M=2.01~M_\odot$ for $m_0 = 700~$MeV and $\alpha b_0=370~$MeV. The~red regions show the phase, where the chiral symmetry is broken, in~the blue regions chiral symmetry is restored. The~phases are separated with the dashed~lines.}
    \label{fig:profiles}
  \end{figure}

In Figure~\ref{fig:m_panel}, we show \mbox{mass vs. central net-baryon density} relations obtained for different values of the chirally invariant mass $m_0$. What is evident is that increasing the value of $m_0$ strengthens the chiral phase transition. This is seen twofold, as~a shrinking of the coexistence phases  and as more abrupt flattening of chirally restored branches. For~a larger $m_0$, the transition becomes strong enough to produce disconnected branches (see, e.g.,~the red, solid line in the bottom right panel of Figure~\ref{fig:m_panel}). These, in~turn, cause the maximal mass of the sequences to decrease with increasing value of $m_0$. Eventually, the~equations of state become not stiff enough to reach the $2~M_\odot$ constraint. We note that such small maximal masses are result of the additional six-point interaction term considered in the thermodynamic potential of the hybrid QMN model (see Equation~\eqref{eq:potentials}). For~$m_0=700~$MeV (Figure~\ref{fig:m_panel}, top), the~most favorable parameterizations are $\alpha b_0=350-370~$MeV, while for $m_0=800~$MeV (Figure~\ref{fig:m_panel}, bottom) none of the EoSs is stiff enough. In~Table~\ref{tab:max_mass}, we show the values of the maximal masses of neutron star and corresponding radii obtained in each parameterization. In~Figure~\ref{fig:profiles}, we show the radial profiles of energy density (Figure~\ref{fig:profiles}, top) and pressure (Figure~\ref{fig:profiles}, bottom), for~a $2.01~M_\odot$ neutron star, calculated for $m_0=700~$MeV and $\alpha b_0=370~$MeV. The~chiral phase transition happens at roughly $7.4~$km from the center of the star and is reflected in a jump in energy~density.

  \begin{table}[H]\centering
    \caption{Maximal neutron-star masses in units of $M_\odot$ and corresponding radius in km (separated by comma) for different values of $m_0$ and $\alpha b_0$ parameters.}
    \label{tab:max_mass}
  \begin{tabular}{ccccc}
    \toprule
                & \multicolumn{4}{c}{\boldmath{$\alpha b_0~$} \textbf{[MeV]}}         \\ \midrule
    \boldmath{$m_0$}~\textbf{[MeV]} & \textbf{350}       & \textbf{370}      & \textbf{400}      & \textbf{450}  \\ \midrule 
    700         & $2.10,~12.11$ & $2.05,~11.91$ & $2.01,~11.81$ & $1.96,~11.92$ \\  
    800         & $1.95,~11.64$ & $1.88,~11.29$ & $1.83,~11.22$ & $1.79,~11.25$ \\ \bottomrule
    \end{tabular}
  
  \end{table}

  The end points of the mass--radius relation correspond to the onset of quark d.o.f. in each parameterization. This leads to the conclusion that the hadronic matter is not stiff enough to fulfill the two-solar-mass constraint. In~general, a~possible resolution to this problem could be another phase transition. This is the case in the hybrid QMN model, which features sequential chiral and deconfinement phase transitions. However, in~the current model setup, the~equation of state in the deconfined phase is not stiff enough to sustain the gravitational collapse and the branches become immediately unstable. This is because quarks are not coupled with the vector field leading to a repulsive force. On~the other hand, it is known that repulsive interactions tend to stiffen the equation of state. Hence, an~additional repulsive force in the quark sector could possibly make the branch stiff enough in order to reach the $2~M_\odot$ constraint, and~an additional family of stable hybrid compact stars would appear, with~the possibility for the \mbox{high-mass} twin scenario advocated by other effective models~\cite{Alvarez-Castillo:2017qki, Ayriyan:2017nby, Kaltenborn:2017hus}. 

  We note that the obtained \mbox{mass--radius} relations stay in good agreement with the low-mass constraints derived from the recent neutron-star merger GW170817 for the low- and high-mass posteriors~\cite{Abbott:2018exr}. In~  Figure~\ref{fig:m_panel} (right), they are shown as green and purple regions, respectively.


\unskip

\unskip
  
\section{Isospin-Symmetric Phase~Diagram}
\label{sec:qcd_phase_diagram}

\begin{figure}[H]
  \centering
    \includegraphics[width=.49\linewidth]{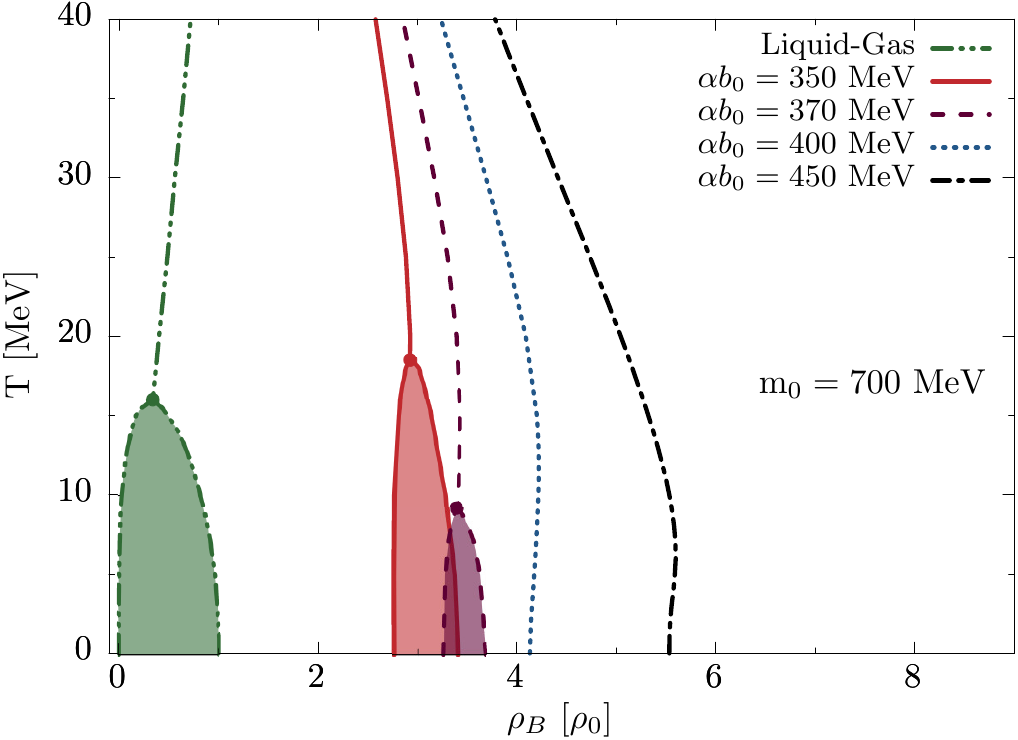}
    \includegraphics[width=.49\linewidth]{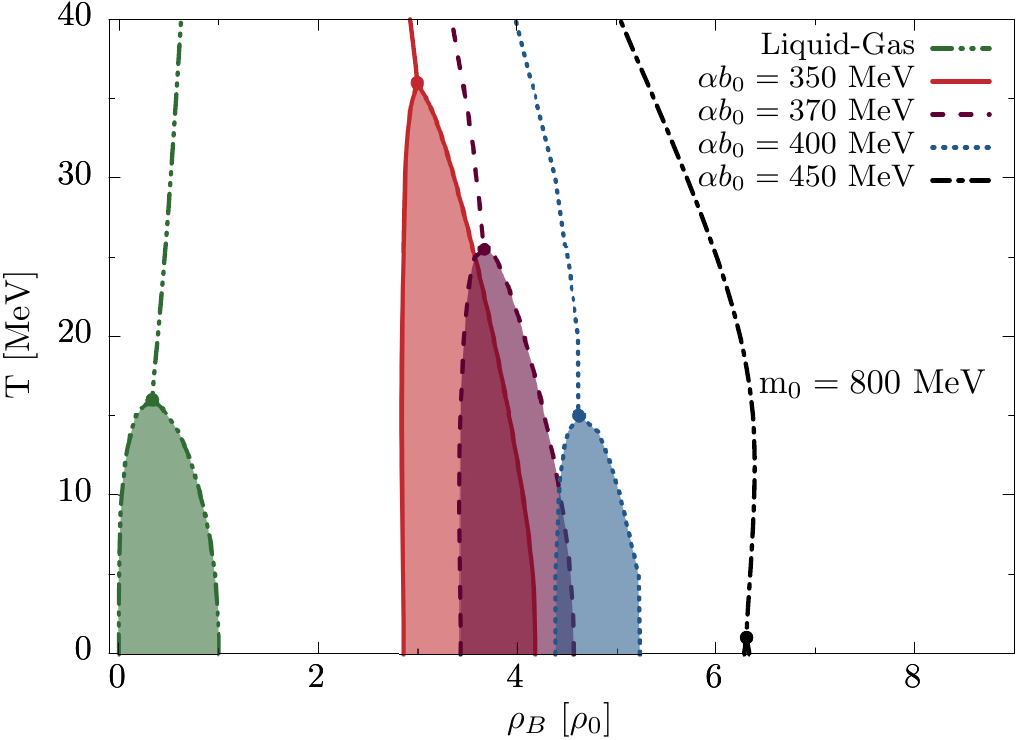}
    \caption{The Low-temperature part of phase diagram in the $(T,\rho_B)$-plane for isospin-symmetric matter obtained in the hybrid QMN model for $m_0=700~$MeV (\textbf{left)} and $m_0=800~$MeV (\textbf{right}). The~curves indicate phase boundaries and the colored areas correspond to the density jump associated with the first-order phase transition. The~green \mbox{dashed-doubly-dotted} curve corresponds to the liquid--gas phase transition common for all $\alpha b_0$. The~circles indicate critical points on the transition lines above which the first-order transition turns into a crossover. For~$m_0=700~$MeV, no critical point is shown for the cases with $\alpha b_0=400~$MeV and $\alpha b_0=450$~MeV, where the chiral phase transition is a smooth crossover at all~temperatures.}
    \label{fig:phase_diag}
  \end{figure}

  The observational neutron-star data provide useful constraints on the structure of strongly interacting matter. Furthermore, they may constrain the phase diagram of isospin-symmetric QCD matter, which is of major relevance for the heavy-ion physics. In~Figure~\ref{fig:phase_diag}, we show the low-temperature part of the isospin-symmetric phase diagram obtained in the model in the \mbox{$(T,\rho_B)$-plane} for \mbox{$m_0=700~$MeV~(Figure~\ref{fig:phase_diag}, left)} and \mbox{$m_0=800~$MeV~(Figure~\ref{fig:phase_diag}, right)}. The~liquid--gas phase transition (green, \mbox{dashed-doubly-dotted} line) is common for both values of $m_0$ by construction of the hybrid QMN model~\cite{Benic:2015pia, Marczenko:2017huu}. Its critical point shows up at around $T=16~$MeV, above~which it turns into crossover. A similar phase structure is developed for the chiral phase transition for low values of $\alpha b_0$. For~$m_0=700~$MeV, the~critical points appear around $T=19, ~9~$MeV for $\alpha b_0=350,~370~$MeV, respectively. On~the other hand, for~$\alpha b_0=400,~450~$MeV, the~chiral transition proceeds as a smooth crossover at all temperatures. For~$m_0=800~$MeV, the~critical points are developed at $T=36,~26,15,1~$MeV for $\alpha b_0=350,~370,~400,~450~$MeV, respectively. Higher values of the temperatures for the critical points are essentially a result of much stronger chiral phase transition at zero temperature. We note that the most favorable parameterizations, i.e.,~for smaller values of $m_0$, yield rather low temperature for the critical point of the chiral phase~transition.

\section{Conclusions}
\label{sec:conclusions}

  In this work, we investigated the hybrid QMN model for the equation of state of dense matter under neutron-star conditions and the phenomenology of compact stars. In~particular, we focused on the implications of including six-point scalar interaction and studied the consequences of the realization of the chiral symmetry restoration within the hadronic~phase.

  We   found that the apparent softening of the EoS results in mass--radius relations with maximal mass at tension with the $2~M_\odot$ constraint within the hadronic branch, especially if the new PSR~J0740+6620 with $2.17~M_\odot$~\cite{Cromartie:2019kug} is considered. We have shown that parameterizations of the model which yield large maximal mass (i.e., for smaller value of $m_0$)  suggest rather low value of the temperature for the critical end point of the first-order chiral phase transition in the phase diagram, which may also be absent. In~view of this, if~would interesting to establish a constraint on the chirally invariant mass $m_0$. Since the hybrid QMN model features sequential chiral and deconfinement phase transitions, one possible resolution to this could be the onset of quark d.o.f. Such a scenario would be even further supported in view of the recent formulation of the three-flavor parity doubling~\cite{Steinheimer:2011ea, Sasaki:2017glk} and further lattice QCD studies~\cite{Aarts:2018glk}, where it was found that, to a large extent, the phenomenon occurs also in the hyperon channels. In~general, the~inclusion of heavier flavors is known to soften the equation of state and additional repulsive forces are needed to comply with the $2~M_\odot$ constraint. Additional stiffness from the quark side would play a role, which is not included in the current study. Work in this direction is in progress and the results will be reported~elsewhere.

\authorcontributions{Conceptualization, M.M. and D.B.; Data curation, M.M.; Formal analysis, M.M.; Funding acquisition, M.M., D.B. and~K.R.; Investigation, M.M.; Methodology, M.M.; Project administration, M.M.; Resources, M.M.; Software, M.M.; Supervision, D.B., K.R. and~C.S.; Validation, M.M.; Visualization, M.M.; Writing---original draft, M.M.; and Writing---review and editing, M.M., D.B., K.R. and~C.S.}

\funding{This work was partly supported by the Polish National Science Center (NCN), under~Maestro Grant No. DEC-2013/10/A/ST2/00106 (K.R. and C.S.), Opus Grant No. 2018/31/B/ST2/01663 (K. R. and C.S.), and~Preludium Grant No. UMO-2017/27/N/ST2/01973 (M.M.). D.B. is grateful for support from the Russian Science Foundation under Contract No. 17-12-01427. We acknowledge the COST Actions CA15213 ``THOR'' and CA16214 ``PHAROS'' for supporting networking activities.}

\acknowledgments{M.M. acknowledges fruitful discussions with N.-U.~F.~Bastian~and~T.~Fischer. M.M. would like to thank the organizing committee of the {\it Compact Stars in the QCD Phase Diagram~VII} for the chance to present his work and for the great atmosphere during the conference.}

\conflictsofinterest{The authors declare no conflict of~interest.} 
  \reftitle{References}


\begin{thebibliography}{999}


\bibitem[Bazavov  {et~al.}(2014)Bazavov et~al.]{Bazavov:2014pvz}
Bazavov, A.; Bhattacharya, T.; DeTar, C.; Ding, H.-T.; Gottlieb, S.; Gupta, R.; Hegde, P.; Heller, U.; Karsch, F.; Laermann, E.; et al.
\newblock {Equation of state in ( 2+1 )-flavor QCD}.
\newblock {\em Phys. Rev. D} {\bf 2014}, {\em 90},~094503.

\bibitem[Lindblom(1998)]{Lindblom:1998dp}
Lindblom, L.
\newblock {Phase transitions and the mass radius curves of relativistic stars}.
\newblock {\em Phys. Rev. D} {\bf 1998}, {\em 58},~024008.

\bibitem[Tolman(1939)]{Tolman:1939jz}
Tolman, R.C.
\newblock {Static solutions of Einstein's field equations for spheres of
  fluid}.
\newblock {\em Phys. Rev.} {\bf 1939}, {\em 55},~364--373.

\bibitem[Oppenheimer and Volkoff(1939)]{Oppenheimer:1939ne}
Oppenheimer, J.R.; Volkoff, G.M.
\newblock {On Massive neutron cores}.
\newblock {\em Phys. Rev.} {\bf 1939}, {\em 55},~374--381.

\bibitem[Steiner  {et~al.}(2010)Steiner, Lattimer, and Brown]{Steiner:2010fz}
Steiner, A.W.; Lattimer, J.M.; Brown, E.F.
\newblock {The Equation of State from Observed Masses and Radii of Neutron
  Stars}.
\newblock {\em Astrophys. J.} {\bf 2010}, {\em 722},~33--54.

\bibitem[Steiner  {et~al.}(2013)Steiner, Lattimer, and Brown]{Steiner:2012xt}
Steiner, A.W.; Lattimer, J.M.; Brown, E.F.
\newblock {The Neutron Star Mass-Radius Relation and the Equation of State of
  Dense Matter}.
\newblock {\em Astrophys. J.} {\bf 2013}, {\em 765},~L5.

\bibitem[Alvarez-Castillo  {et~al.}(2016)Alvarez-Castillo, Ayriyan, Benic,
  Blaschke, Grigorian, and Typel]{Alvarez-Castillo:2016oln}
Alvarez-Castillo, D.; Ayriyan, A.; Benic, S.; Blaschke, D.; Grigorian, H.;
  Typel, S.
\newblock {New class of hybrid EoS and Bayesian M-R data analysis}.
\newblock {\em Eur. Phys. J. A} {\bf 2016}, {\em 52},~69.

\bibitem[Cromartie  {et~al.}(2019)Cromartie et~al.]{Cromartie:2019kug}
Cromartie, H.T.; Fonseca, E.; Ransom, S.M.; Demorest, P.B.; Arzoumanian, Z.; Blumer, H.; Brook, P.R.; DeCesar, M.E.; Dolch, T.; Ellis, J.A.; et al.
\newblock {A very massive neutron star: Relativistic Shapiro delay measurements
  of PSR J0740+6620.} \emph{arXiv}  {\bf 2019}, arXiv:1904.06759.

\bibitem[Abbott  {et~al.}(2018)Abbott et~al.]{Abbott:2018exr}
Abbott, B.P. et al. [LIGO Scientific Collaboration and Virgo Collaboration]
\newblock {GW170817: Measurements of neutron star radii and equation of state}.
\newblock {\em Phys. Rev. Lett.} {\bf 2018}, {\em 121},~161101.

\bibitem[Alford  {et~al.}(2013)Alford, Han, and Prakash]{Alford:2013aca}
Alford, M.G.; Han, S.; Prakash, M.
\newblock {Generic conditions for stable hybrid stars}.
\newblock {\em Phys. Rev. D} {\bf 2013}, {\em 88},~083013.

\bibitem[Blaschke  {et~al.}(2013)Blaschke, Alvarez-Castillo, and
  Benic]{Blaschke:2013ana}
Blaschke, D.; Alvarez-Castillo, D.E.; Benic, S.
\newblock {Mass-radius constraints for compact stars and a critical endpoint}.
\newblock {\em arXiv} {\bf 2013}, arXiv:1310.3803.

\bibitem[Benic  {et~al.}(2015)Benic, Blaschke, Alvarez-Castillo, Fischer, and
  Typel]{Benic:2014jia}
Benic, S.; Blaschke, D.; Alvarez-Castillo, D.E.; Fischer, T.; Typel, S.
\newblock {A new quark-hadron hybrid equation of state for astrophysics - I.
  High-mass twin compact stars}.
\newblock {\em Astron. Astrophys.} {\bf 2015}, {\em 577},~A40.

\bibitem[Alvarez-Castillo  {et~al.}(2016)Alvarez-Castillo, Benic, Blaschke,
  Han, and Typel]{Alvarez-Castillo:2016wqj}
Alvarez-Castillo, D.; Benic, S.; Blaschke, D.; Han, S.; Typel, S.
\newblock {Neutron star mass limit at $2M_\odot$ supports the existence of a
  CEP}.
\newblock {\em Eur. Phys. J. A} {\bf 2016}, {\em 52},~232.

\bibitem[Marczenko  {et~al.}(2018)Marczenko, Blaschke, Redlich, and
  Sasaki]{Marczenko:2018jui}
Marczenko, M.; Blaschke, D.; Redlich, K.; Sasaki, C.
\newblock {Chiral symmetry restoration by parity doubling and the structure of
  neutron stars}.
\newblock {\em Phys. Rev. D} {\bf 2018}, {\em 98},~103021.

\bibitem[Benic  {et~al.}(2015)Benic, Mishustin, and Sasaki]{Benic:2015pia}
Benic, S.; Mishustin, I.; Sasaki, C.
\newblock {Effective model for the QCD phase transitions at finite baryon
  density}.
\newblock {\em Phys. Rev. D} {\bf 2015}, {\em 91},~125034.

\bibitem[Marczenko and Sasaki(2018)]{Marczenko:2017huu}
Marczenko, M.; Sasaki, C.
\newblock {Net-baryon number fluctuations in the Hybrid Quark-Meson-Nucleon
  model at finite density}.
\newblock {\em Phys. Rev. D} {\bf 2018}, {\em 97},~036011.

\bibitem[Detar and Kunihiro(1989)]{Detar:1988kn}
Detar, C.E.; Kunihiro, T.
\newblock {Linear $\sigma$ Model With Parity Doubling}.
\newblock {\em Phys. Rev. D} {\bf 1989}, {\em 39},~2805--2808.

\bibitem[Jido  {et~al.}(2000)Jido, Hatsuda, and Kunihiro]{Jido:1999hd}
Jido, D.; Hatsuda, T.; Kunihiro, T.
\newblock {Chiral symmetry realization for even parity and odd parity baryon
  resonances}.
\newblock {\em Phys. Rev. Lett.} {\bf 2000}, {\em 84},~3252--3255.

\bibitem[Jido  {et~al.}(2001)Jido, Oka, and Hosaka]{Jido:2001nt}
Jido, D.; Oka, M.; Hosaka, A.
\newblock {Chiral symmetry of baryons}.
\newblock {\em Prog. Theor. Phys.} {\bf 2001}, {\em 106},~873--908.

\bibitem[Patrignani  {et~al.}(2016)Patrignani et~al.]{Patrignani:2016xqp}
 Patrignani, C. et al. [Particle Data Group]
\newblock {Review of Particle Physics}.
\newblock {\em Chin. Phys. C} {\bf 2016}, {\em 40},~100001.

\bibitem[Roberts  {et~al.}(2010)Roberts, Roberts, Bashir, Gutierrez-Guerrero,
  and Tandy]{Roberts:2010rn}
Roberts, H.L.L.; Roberts, C.D.; Bashir, A.; Gutierrez-Guerrero, L.X.; Tandy,
  P.C.
\newblock {Abelian anomaly and neutral pion production}.
\newblock {\em Phys. Rev. C} {\bf 2010}, {\em 82},~065202.
\newblock

\bibitem[Roberts  {et~al.}(2011)Roberts, Bashir, Gutierrez-Guerrero, Roberts,
  and Wilson]{Roberts:2011wy}
Roberts, H.L.L.; Bashir, A.; Gutierrez-Guerrero, L.X.; Roberts, C.D.; Wilson,
  D.J.
\newblock {pi- and rho-mesons, and their diquark partners, from a contact
  interaction}.
\newblock {\em Phys. Rev. C} {\bf 2011}, {\em 83},~065206.
\newblock

\bibitem[Lattimer and Lim(2013)]{Lattimer:2012xj}
Lattimer, J.M.; Lim, Y.
\newblock {Constraining the Symmetry Parameters of the Nuclear Interaction}.
\newblock {\em Astrophys. J.} {\bf 2013}, {\em 771},~51.

\bibitem[Motohiro  {et~al.}(2015)Motohiro, Kim, and Harada]{Motohiro:2015taa}
Motohiro, Y.; Kim, Y.; Harada, M.
\newblock {Asymmetric nuclear matter in a parity doublet model with hidden
  local symmetry}.
\newblock {\em Phys. Rev. C} {\bf 2015}, {\em 92},~025201.

\bibitem[Zschiesche  {et~al.}(2007)Zschiesche, Tolos, Schaffner-Bielich, and
  Pisarski]{Zschiesche:2006zj}
Zschiesche, D.; Tolos, L.; Schaffner-Bielich, J.; Pisarski, R.D.
\newblock {Cold, dense nuclear matter in a SU(2) parity doublet model}.
\newblock {\em Phys. Rev. C} {\bf 2007}, {\em 75},~055202.

\bibitem[Aarts  {et~al.}(2017)Aarts, Allton, De~Boni, Hands, Jäger, Praki,
  and Skullerud]{Aarts:2017rrl}
Aarts, G.; Allton, C.; De~Boni, D.; Hands, S.; Jäger, B.; Praki, C.;
  Skullerud, J.I.
\newblock {Light baryons below and above the deconfinement transition: Medium
  effects and parity doubling}.
\newblock {\em J. High Energy Phys.} {\bf 2017}, {\em 2017},~034.

\bibitem[Aarts  {et~al.}(2019)Aarts, Allton, De~Boni, and
  Jäger]{Aarts:2018glk}
Aarts, G.; Allton, C.; De~Boni, D.; Jäger, B.
\newblock {Hyperons in thermal QCD: A lattice view}.
\newblock {\em Phys. Rev. D} {\bf 2019}, {\em 99},~074503.

\bibitem[Collins  {et~al.}(1977)Collins, Duncan, and
  Joglekar]{Collins:1976yq}
Collins, J.C.; Duncan, A.; Joglekar, S.D.
\newblock {Trace and Dilatation Anomalies in Gauge Theories}.
\newblock {\em Phys. Rev. D} {\bf 1977}, {\em 16},~438--449.
\newblock

\bibitem[Bardeen  {et~al.}(1986)Bardeen, Leung, and Love]{Bardeen:1985sm}
Bardeen, W.A.; Leung, C.N.; Love, S.T.
\newblock {The Dilaton and Chiral Symmetry Breaking}.
\newblock {\em Phys. Rev. Lett.} {\bf 1986}, {\em 56},~1230--1233.
\newblock

\bibitem[Nielsen(1977)]{Nielsen:1977sy}
Nielsen, N.K.
\newblock {The Energy Momentum Tensor in a Nonabelian Quark Gluon Theory}.
\newblock {\em Nucl. Phys. B} {\bf 1977}, {\em 120},~212--220.
\newblock

\bibitem[Alford  {et~al.}(2015)Alford, Burgio, Han, Taranto, and
  Zappalà]{Alford:2015dpa}
Alford, M.G.; Burgio, G.F.; Han, S.; Taranto, G.; Zappalà, D.
\newblock {Constraining and applying a generic high-density equation of state}.
\newblock {\em Phys. Rev. D} {\bf 2015}, {\em 92},~083002,
\newblock

\bibitem[Antoniadis  {et~al.}(2013)Antoniadis et~al.]{Antoniadis:2013pzd}
Antoniadis, J.; Freire, P.C.; Wex, N.; Tauris, T.M.; Lynch, R.S.; van Kerkwijk, M.H.; Kramer, M.; Bassa, C.; Dhillon, V.S.; Driebe, T.; et al.
\newblock {A Massive Pulsar in a Compact Relativistic Binary}.
\newblock {\em Science} {\bf 2013}, {\em 340},~6131.

\bibitem[Alvarez-Castillo and Blaschke(2017)]{Alvarez-Castillo:2017qki}
Alvarez-Castillo, D.E.; Blaschke, D.B.
\newblock {High-mass twin stars with a multipolytrope equation of state}.
\newblock {\em Phys. Rev. C} {\bf 2017}, {\em 96},~045809.

\bibitem[Ayriyan  {et~al.}(2018)Ayriyan, Bastian, Blaschke, Grigorian,
  Maslov, and Voskresensky]{Ayriyan:2017nby}
Ayriyan, A.; Bastian, N.U.; Blaschke, D.; Grigorian, H.; Maslov, K.;
  Voskresensky, D.N.
\newblock {Robustness of third family solutions for hybrid stars against mixed
  phase effects}.
\newblock {\em Phys. Rev. C} {\bf 2018}, {\em 97},~045802.

\bibitem[Kaltenborn  {et~al.}(2017)Kaltenborn, Bastian, and
  Blaschke]{Kaltenborn:2017hus}
Kaltenborn, M.A.R.; Bastian, N.U.F.; Blaschke, D.B.
\newblock {Quark-nuclear hybrid star equation of state with excluded volume
  effects}.
\newblock {\em Phys. Rev. D} {\bf 2017}, {\em 96},~056024.

\bibitem[Steinheimer  {et~al.}(2011)Steinheimer, Schramm, and
  Stocker]{Steinheimer:2011ea}
Steinheimer, J.; Schramm, S.; Stocker, H.
\newblock {The hadronic SU(3) Parity Doublet Model for Dense Matter, its
  extension to quarks and the strange equation of state}.
\newblock {\em Phys. Rev. C} {\bf 2011}, {\em 84},~045208.

\bibitem[Sasaki(2018)]{Sasaki:2017glk}
Sasaki, C.
\newblock {Parity doubling of baryons in a chiral approach with three flavors}.
\newblock {\em Nucl. Phys. A} {\bf 2018}, {\em 970},~388--397.

\end{thebibliography}
%

\end{document}